# Portfolio selection problems in practice: a comparison between linear and quadratic optimization models

Francesco Cesarone*, Andrea Scozzari‡, Fabio Tardella§

*Università degli Studi Roma Tre - Dipartimento di Economia

fcesarone@uniroma3.it

‡UNISU - Università Telematica delle Scienze Umane "Niccolò Cusano"

Facoltà di Economia

andrea.scozzari@unisu.it

§Università di Roma "La Sapienza"

Dipartimento di Metodi e Modelli per l'Economia, il Territorio e la Finanza

fabio.tardella@uniroma1.it

ABSTRACT

Several portfolio selection models take into account practical limitations on the number of assets to include and on their weights in the portfolio. We present here a study of the *Limited Asset* Markowitz (LAM), of the *Limited Asset* Mean Absolute Deviation (LAMAD) and of the *Limited Asset* Conditional Value-at-Risk (LACVaR) models, where the assets are limited with the introduction of quantity and cardinality constraints.

We propose a completely new approach for solving the LAM model, based on reformulation as a Standard Quadratic Program and on some recent theoretical results. With this approach we obtain optimal solutions both for some well-known financial data sets used by several other authors, and for some unsolved large size portfolio problems. We also test our method on five new data sets involving real-world capital market indices from major stock markets. Our computational experience shows that, rather unexpectedly, it is easier to solve the quadratic LAM model with our algorithm, than to solve the linear LACVaR and LAMAD models with CPLEX, one of the best commercial codes for mixed integer linear programming (MILP) problems.

Finally, on the new data sets we have also compared, using out-of-sample analysis, the performance of the portfolios obtained by the *Limited Asset* models with the performance provided by the unconstrained models and with that of the official capital market indices.

# 1 Introduction

The classical Mean-Variance (MV) portfolio selection model of Markowitz [40, 41, 42] has been widely recognized as one of the cornerstones of modern portfolio theory. However, its success has inevitably drawn many criticisms and proposals of alternative or more refined models (see, e.g., [19, 28, 32, 43, 44, 46, 47] and references therein).

Among the many refinements that have been proposed to make the Markowitz model more realistic, we analyze in this paper the one that limits the number of assets to be held in an efficient portfolio (*cardinality constraint*), and also the one that prescribes lower and upper bounds on the fraction of the capital invested in each asset (*quantity constraints*). These requirements come from real-world practice, where the administration of a portfolio made up of a large number of assets, possibly with very small holdings for some of them, is clearly not desirable because of transactions costs, minimum lot sizes, complexity of management, or policy of the asset management companies. We call *Limited Asset* Markowitz (LAM) model the Markowitz model with the above restrictions. Because of its practical relevance, this model (often called *cardinality constrained Markowitz model*), and some variations thereof, have been fairly intensively studied in the last decade, especially from the computational viewpoint [4, 9, 10, 14, 16, 17, 18, 20, 21, 24, 26, 35, 39, 45, 50, 52, 55]. In these studies it appears that the computational complexity for the solution of the LAM model is much greater than the one required by the classical Markowitz model or by several other of its refinements. Indeed, the standard Markowitz model is routinely solved for markets with thousands of assets. This practical difference in computational complexity is also theoretically justified by the fact that the classical Markowitz model is a convex quadratic programming problem that has a polynomial worst-case complexity bound, while the LAM model is usually modeled by adding binary variables, thus becoming a mixed integer quadratic programming (MIQP) problem, that falls into the class of considerably more difficult NP-hard problems (see, e.g., [10, 52]).

We remark that some attempts have been made to construct simpler portfolio selection models, for instance, by linearizing the quadratic objective function. These approaches involve either the approximation or the decomposition of the covariance matrix (see, e.g., [44]). Some researchers have also introduced alternative risk measures for portfolio planning. In many cases these measures are linear, leading to a corresponding simplification from the computational viewpoint. Konno [30] introduced the mean absolute deviation (MAD) model. This return-risk model together with the return-standard deviation model are among the most important portfolio models, that use dispersion measures. Speranza [53] considered the downside mean semi-deviation, i.e., the mean absolute value of negative deviations. She also proved that it is always equal to half of the mean absolute deviation from the mean. Roy [49] laid the basis for the development of downside risk measures. The objective in portfolio selection models with downside risk measures is the maximization of the probability that the portfolio return is above a certain minimal acceptable level. Markowitz [41] proposed the semi-variance as an alternative risk measure, but he suggested that the use of variance is computationally more tractable and reveals the same information. In the 1970s, many papers (see, e.g., [6, 22]) provided a natural generalization of semi-variance with the lower partial moment risk measure, whose justification proceeds from the observation that an investor's true risk is the downside risk. Young [59] developed a minimax approach for the portfolio management, measuring risk as the minimum return (maximum loss) that the portfolio would have achieved over all of the past observation periods. In 1994 JP Morgan proposed what is now probably the most famous downside risk measure: Value-at-Risk (VaR, see [46]). This is a very important risk management tool in the financial industry, but with some drawbacks: the VaR optimization problem is not convex, and VaR is not sub-additive, i.e., it does not express the benefits of diversification. Sub-additivity is one of a set of desirable properties that define a coherent risk measure (see the results described by Artzner *et al.* [5] for downside risk measures, and by Rockafellar *et al.* [48] for dispersion measures). One of the most important coherent risk measures is Conditional Value-at-Risk (CVaR, see [47]). Some researchers call it Expected Shortfall [2]. Enhanced CVaR measures has been suggested by Mansini *et al.* [36]. Acerbi [1] proposed a spectral risk measure which involves a weighted average of the quantiles of the loss. CVaR is a special case of spectral risk measure. Although much work has been done on risk measures and mean-risk models, the question of which risk measure is most appropriate is still the subject of much debate.

In this paper we compare the LAM model with the CVaR and MAD portfolio models under additional constraints that limit the number of assets to be held in a portfolio and prescribe lower and upper bounds on the fraction of the capital invested in each asset. We call them *Limited Asset* CVaR (LACVaR) model and *Limited Asset* MAD (LAMAD) model, respectively. These two models are usually formulated as Mixed Integer Linear Programming models (MILP) and fall into the class of NP-hard problems too.

The LAM model is solved here with a completely new approach that is based on a reformulation as a Standard Quadratic Programming problem and exploits recent theoretical results for Quadratic Programming by Tardella [57, 58] and by Scozzari and Tardella [51]. Our method is able to solve to optimality the well-known five benchmark problems described in [14] and publicly available in Beasley's OR Library [7], whose optimal solution has been reported only very recently by Di Gaspero *et al.* [18]. In addition to these five problems, we report solutions of much larger and unsolved real-world problems, one with around 500 assets and two with more than 2000 assets also taken from the OR-Library [13]. We provide some computational results comparing our solution method with the exact MIQP solver implemented in CPLEX 11.0. We also test our method on five new data sets involving real-world capital market indices from major stock markets. Our experimental analysis shows that the practical computational complexity for most exact algorithms for the LAM model seems to be related not only to the number of variables but also to the number of assets with positive weight in the solution of the unconstrained Markowitz model.

An important issue highlighted in this study is that, rather unexpectedly, it easier to solve the quadratic LAM model with our algorithm, than to solve the linear LACVaR and LAMAD models with CPLEX, one of the best commercial codes for MILP problems.

Finally, on the new data sets we have compared, using out-of-sample analysis, the performance of the portfolios obtained by the *Limited Asset* models with the performance provided by the unconstrained models and with that of the official capital market indices.

We made our data sets and the solutions that we found publicly available for use by other researchers in this field.

The paper is organized as follows: Section 2 introduces the three *Limited Asset* models considered in this study along with a review of the methods used to solve them. In Sections 3 and 4 we present our new approach to solve the LAM model based on a reduction to a Standard Quadratic Programming problem. Section 5 reports some computational results showing that the quadratic LAM model can be solved more efficiently than the linear LACVaR and LAMAD models. In Section 6 we present a comparison among the ex-post performances of the portfolios obtained by the models.

## 2 Limited Asset Models

### 2.1 The Limited Asset Markowitz Model

The classical Mean-Variance (MV) portfolio optimization model introduced by Markowitz aims at determining the fractions $x_i$ of a given capital to be invested in each asset $i$ belonging to a predetermined set or market so as to minimize the risk of the return of the whole portfolio, identified with its variance, while restricting the expected return of the portfolio to attain a specified value.

More precisely, we assume that $n$ assets are available, and we denote by $\mu_i$ the expected return of asset $i$, and by $\sigma_{ij}$ the covariance of returns of asset $i$ and asset $j$. We also denote by $\rho$ the required level of return for the portfolio. The classical MV model is:

$$\begin{aligned}
\text{Min} \quad & \sum_{i=1}^{n}\sum_{j=1}^{n} \sigma_{ij} x_i x_j \\
\text{st} \quad & \\
& \sum_{i=1}^{n} \mu_i x_i = \rho \\
& \sum_{i=1}^{n} x_i = 1 \\
& x_i \geq 0 \qquad i = 1,\ldots,n
\end{aligned} \qquad (1)$$

This is a convex quadratic programming problem which can be solved by a number of efficient algorithms with a moderate computational effort even for large instances. We denote by $\phi(\rho)$ the optimal value of (1) as a function of $\rho$. Let $\rho_{min}$ denote the value of $\sum_{i=1}^{n} \mu_i x_i$ at an optimal solution of the problem obtained by deleting the first constraint in (1), and let $\rho_{max} = \max\{\mu_1,\ldots,\mu_n\}$. Then the graph of $\phi(\rho)$ on the interval $[\rho_{min}, \rho_{max}]$ coincides with the set of all non-dominated (or efficient) portfolios (*efficient frontier*), and is usually approximated by solving (1) for several (equally spaced) values of $\rho$ in $[\rho_{min}, \rho_{max}]$.

**Proposition 1** *The convexity of (1) implies that, for $\rho \geq \rho_{min}$, the function $\phi(\rho)$ is increasing and convex.*

**Proof.** Let $\rho_0, \rho_1 \in [\rho_{min}, \rho_{max}]$ and let $x_0, x_1$ be two corresponding solutions of (1). Then $x_\lambda = (1-\lambda)x_0 + \lambda x_1$ is a feasible solution to (1) for $\rho_\lambda = (1-\lambda)\rho_0 + \lambda \rho_1$, for any $\lambda \in [0,1]$ (due to the linearity of expected return). The convexity of $\phi(\rho)$ follows from the convexity of the variance $\sigma^2(x) = x'\Sigma x$:

$$\phi(\rho_\lambda) \leq \sigma^2(x_\lambda) = \sigma^2((1-\lambda)x_0 + \lambda x_1) \leq (1-\lambda)\sigma^2(x_0) + \lambda\sigma^2(x_1) = (1-\lambda)\phi(\rho_0) + \lambda\phi(\rho_1)$$

To prove isotonicity of $\phi$, take any $\rho_0 < \rho_1 \in [\rho_{min}, \rho_{max}]$. Then for some $\lambda \in (0,1)$ we have $\rho_0 = \lambda \rho_{min} + (1-\lambda)\rho_1$, so that

$$\phi(\rho_0) \leq \lambda \phi(\rho_{min}) + (1-\lambda)\phi(\rho_1) \leq \lambda \phi(\rho_1) + (1-\lambda)\phi(\rho_1) = \phi(\rho_1),$$

where the last inequality follows from the fact that $\phi(\rho_{min}) \leq \phi(\rho)$ for all $\rho \in [\rho_{min}, \rho_{max}]$, by definition of $\rho_{min}$.

□

From the above proposition we immediately derive the following result:

**Corollary 2** *The solution of (1) does not change if we replace the expected return constraint with $\sum_{i=1}^{n} \mu_i x_i \geq \rho$.*

Note that for every solution $\bar{x}$ of problem (1) the point $(\sum_{i=1}^{n} \mu_i \bar{x}_i, \sum_{i=1}^{n} \sum_{j=1}^{n} \sigma_{ij} \bar{x}_i \bar{x}_j)$ is an efficient point of the multi-objective problem of minimizing risk and maximizing return. Conversely, every such efficient point can be obtained in this manner for some $\rho$.

We now add to the MV model the realistic constraint that no more than $K$ assets should be held in the portfolio (a *cardinality constraint*), and furthermore that the quantity $x_i$ of each asset that is included in the portfolio should be limited within a given interval $[\ell_i, u_i]$ (a *quantity constraint* or *buy-in threshold*). Thus we obtain the following *Limited Asset* Markowitz model:

$$\begin{aligned}
\text{Min} \quad & \sum_{i=1}^{n} \sum_{j=1}^{n} \sigma_{ij} x_i x_j \\
\text{st} \quad & \\
& \sum_{i=1}^{n} \mu_i x_i = \rho \\
& \sum_{i=1}^{n} x_i = 1 \\
& x_i = 0 \text{ or } \ell_i \leq x_i \leq u_i, \quad i = 1, \ldots, n \\
& |supp(x)| \leq K,
\end{aligned} \quad (2)$$

where $supp(x) = \{i : x_i > 0\}$.

Problem (2) is no longer a convex optimization problem because of the non-convexity of its feasible region. As a consequence the optimal value function $\phi_K(\rho)$ of (2) need not be increasing nor convex. Furthermore, $\phi_K(\rho)$ does not any longer coincide with the optimal value function $\phi'_K(\rho)$ of problem (2) where the first constraint is replaced by $\sum_{i=1}^{n} \mu_i x_i \geq \rho$. Indeed, $\phi'_K(\rho) = \phi_K(\rho)$ if and only if the point $(\rho, \phi_K(\rho))$ is on the efficient frontier.

On the other hand, if $\phi'_K(\rho) \neq \phi_K(\rho)$, then there exist no points $(u, v)$ on the efficient frontier with $u = \rho$. We deem important to point out that the non-convexity of problem (2) (due to non-convexity of the feasible region) makes it incorrect to use a convex combination approach to find the efficient frontier for the multi-objective problem of minimizing risk and maximizing return. More precisely, for any $\lambda \in [0, 1]$ and any solution $\bar{x}$ of the problem

$$\begin{aligned}
\text{Min} \quad & \lambda \sum_{i=1}^{n} \sum_{j=1}^{n} \sigma_{ij} x_i x_j - (1-\lambda) \sum_{i=1}^{n} \mu_i x_i \\
\text{st} \quad & \\
& \sum_{i=1}^{n} x_i = 1 \\
& x_i = 0 \text{ or } \ell_i \leq x_i \leq u_i, \quad i = 1, \ldots, n \\
& |supp(x)| \leq K,
\end{aligned} \quad (3)$$

the point $(\sum_{i=1}^{n} \mu_i \bar{x}_i, \sum_{i=1}^{n} \sum_{j=1}^{n} \sigma_{ij} \bar{x}_i \bar{x}_j)$ is an efficient point. However, as observed by Jobst *et al.* [26], not all points on the efficient frontier can be obtained in this way. This difficulty seems to have been overlooked by some authors that used the convex combination approach to build the efficient frontier [20]. We remark that a necessary and sufficient condition for obtaining all the points on the efficient frontier by solving problem (3) for all $\lambda \in [0, 1]$ is that the value function $\phi_K(\rho)$ is convex.

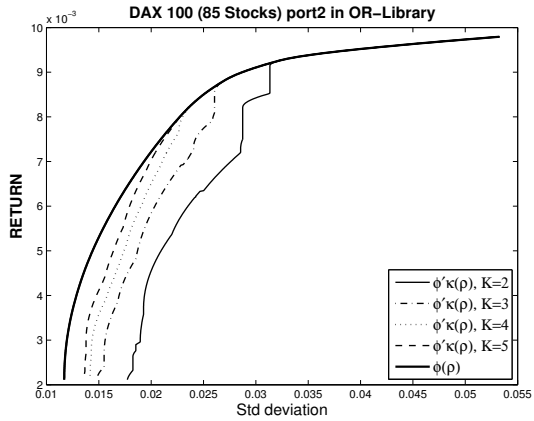

Figure 1: Graphs of $\phi(\rho)$ and $\phi'_K(\rho)$

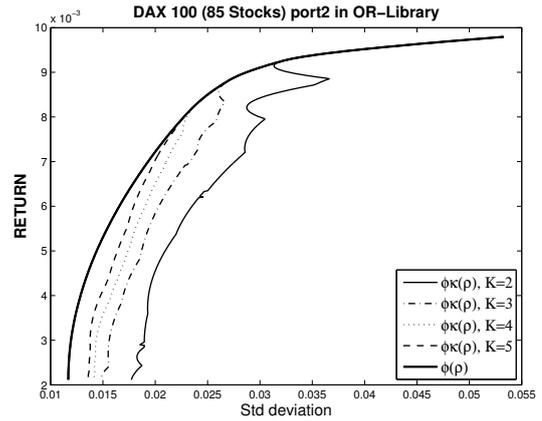

Figure 2: Graphs of $\phi(\rho)$ and $\phi_K(\rho)$

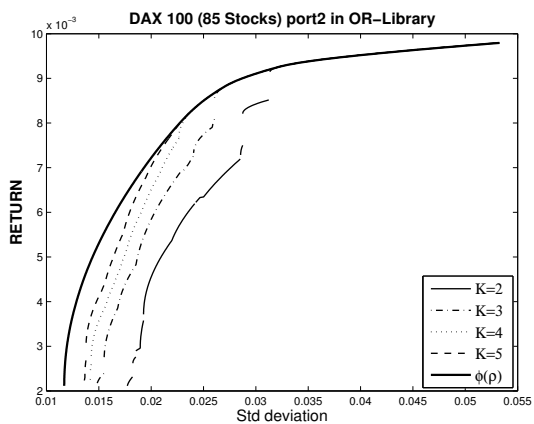

Figure 3: Efficient frontiers

Figs. 1, 2 and 3 illustrate the graphs of $\phi(\rho)$ and $\phi'_K(\rho)$, of $\phi(\rho)$ and $\phi_K(\rho)$, and the efficient frontiers for some values of $K$, in an instance based on real-world data. The lower ($\ell_i$) and upper ($u_i$) bounds are set to 0.01 and 1, respectively. Note that these figures are based on the *exact* optimal solutions to the following problem (4) obtained with the algorithm described in Section 4. Fig. 3 can be compared with Fig. 9 in [14] that is based on *approximate* solutions to (4) found with heuristic algorithms.

As observed by several authors [10, 14, 26], problem (2) can be reformulated as a Mixed Integer Quadratic Program (MIQP) with the addition of $n$ binary variables:

$$\begin{aligned}
\text{Min} \quad & \sum_{i=1}^{n}\sum_{j=1}^{n} \sigma_{ij} x_i x_j \\
\text{st} \quad & \\
& \sum_{i=1}^{n} \mu_i x_i = \rho \\
& \sum_{i=1}^{n} x_i = 1 \\
& \sum_{i=1}^{n} y_i \leq K \\
& \ell_i y_i \leq x_i \leq u_i y_i \quad i = 1, \ldots, n \\
& x_i \geq 0 \quad i = 1, \ldots, n \\
& y_i \in \{0, 1\} \quad i = 1, \ldots, n
\end{aligned} \quad (4)$$

A number of exact approaches have been proposed to solve problem (4). Bienstock [10] proposes a branch-and-cut algorithm and reports good computational results for some real-life problems (not available for comparison). However, his method seems to become extremely slow for small values of $K$. Bertsimas and Shioda [9] extend the algorithm of Bienstock [10] presenting a *tailored procedure*, based on Lemke's pivoting algorithm [34], that takes advantage of the special structure of the problem. They present computational results only on randomly generated data for fairly large values of $K$. A branch-and-bound algorithm for mixed integer nonlinear programs, including portfolio selection problems, is presented in [12]. Li *et al.* [35] propose a convergent Lagrangian method as an exact solution scheme for a problem slightly more general than (4) and they describe some computational results for problems with at most 30 assets. Another Lagrangian relaxation method is proposed in [52] with application to some undisclosed real-life problems with up to 500 assets. Lee and Mitchell [33] develop an interior-point algorithm within a parallel branch-and-bound framework for solving nonlinear mixed-integer programming problems. Preliminary computational results on three randomly generated quadratic portfolio models are reported. Frangioni and Gentile [23] use a method based on perspective cuts to solve randomly generated LAM problems (4) without cardinality constraints involving up to 400 assets. Furthermore, some commercial or free optimization softwares provide tools to solve general MIQPs, and thus (4), although only for problems with few hundreds variables at most.

Since exact methods are able to solve only a fraction of practically useful LAM models, a variety of heuristic procedures have also been proposed for solving (4). Local search techniques are discussed in [50], while Chang *et al.* [14] present three heuristics based upon genetic algorithms, tabu search, and simulated annealing. In [26] two heuristic solution approaches are proposed for problems subject to buy-in threshold, cardinality and roundlot constraints. A hybrid local search algorithm combining principles of simulated annealing and of evolutionary strategies is used in [39] to solve problem (4) in the absence of quantity constraints. Other evolutionary algorithms, combined with local search techniques in order to improve the quality of the solutions, are described in [55, 56]. Fieldsend *et al.* [21] introduce a parallel solution method by extending techniques developed in the multi-objective evolutionary optimization domain. Finally, Di Gaspero *et al.* [18] present a heuristic solver, based on a hybrid technique that combines a local search metaheuristic with a quadratic programming procedure. Experimental results seem to show that their approach is very promising for medium size problems. They also consider *pre-assignment constraints*, which specify a subset of asset that has to be included in the chosen portfolio.

In Section 4 we will show that such constraints actually make the problem easier for the new approach that we propose in this paper.

We should mention that, while several authors experiment their algorithms on undisclosed or randomly generated data, a selection of the cited papers [4, 17, 18, 26, 45, 50, 55, 56] report results obtained on the five real-world data sets introduced in [14] that have been made available by Beasley in his OR-Library [7].

## 2.2 The Limited Asset CVaR and MAD Models

### 2.2.1 CVaR

The *Limited Asset* CVaR (LACVaR) model is similar to the previous one in that it also consists in a risk-return model with realistic constraints, represented by *cardinality* and *quantity constraints*, but with the risk measured by CVaR. The model can be written as follows:

$$
\begin{aligned}
\min \quad & CVaR(x, \epsilon) \\
st \quad & \sum_{i=1}^{n} \mu_i x_i = \rho \\
& \sum_{i=1}^{n} x_i = 1 \\
& x_i = 0 \text{ or } \ell_i \leq x_i \leq u_i, \quad i = 1, \ldots, n \\
& |supp(x)| \leq K,
\end{aligned}
\quad (5)
$$

Fig. 4 shows the optimal solutions of the LACVaR model in the risk-return plane for several data sets and for some values of $K$. As described in [3], problem (5) can be reformulated as a Mixed Integer Linear Program (MILP) with the addition of $n$ binary variables:

$$
\begin{aligned}
\min \quad & \zeta + \frac{1}{\epsilon}\frac{1}{T}\sum_{j=1}^{T} d_j \\
st \quad & \sum_{i=1}^{n} -r_{ij} x_i - d_j - \zeta \leq 0 \quad j = 1, \ldots, T \\
& \sum_{i=1}^{n} \mu_i x_i = \rho \\
& \sum_{i=1}^{n} x_i = 1 \\
& \sum_{i=1}^{n} y_i \leq K \\
& \ell_i y_i \leq x_i \leq u_i y_i && i = 1, \ldots, n \\
& x_i \geq 0 && i = 1, \ldots, n \\
& y_i \in \{0, 1\} && i = 1, \ldots, n \\
& d_j \geq 0 && j = 1, \ldots, T \\
& \zeta \in \mathbb{R}
\end{aligned}
\quad (6)
$$

This is a MILP problem with $n + T + 1$ continuous variables, $n$ binary variables and $T + n + 3$ constraints. In Fig. 4, the bold line represents the frontier in the unconstrained case, while the dashed and the thin lines are the frontiers for some values of $K$. The lower ($\ell_i$) and upper ($u_i$) bounds are set to 0.01 and 1, respectively.

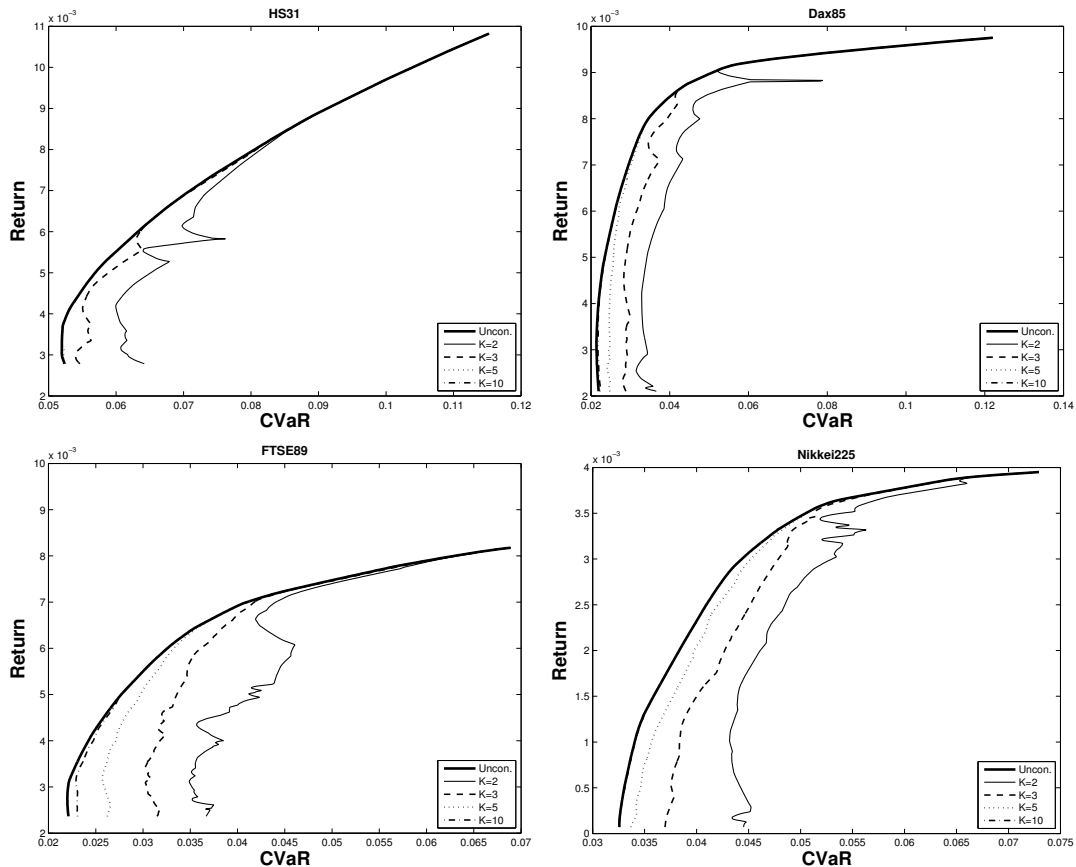

Figure 4: Frontiers of the *Limited Asset* CVaR model

In order to compare the computational time required to solve the LAM and the LACVaR models with respect to the same data sets, we have considered in both models the target return levels $\rho$ used to construct the frontiers for the LAM model. Thus $\rho$ varies in the interval $[\rho_{min}, \rho_{max}]$ for several (equally spaced) values, where $\rho_{min}$ is the value of $\sum_{i=1}^{n} \mu_i x_i$ at an optimal solution of the problem obtained by deleting the constraint on net portfolio mean return in (1), and $\rho_{max} = \max\{\mu_1, \ldots, \mu_n\}$. The optimal solutions and the corresponding frontiers are illustrated in Fig. 4.

### 2.2.2 MAD

The MAD model requires the solution of a simple linear programming problem. However, when we add *cardinality* and *quantity constraints*, it also becomes a Mixed Integer Linear Programming (MILP) for which it is harder to find an optimal solution. This model, called *Limited Asset* MAD (LAMAD) model, can be formulated as follows [3]:

$$\begin{aligned}
\min \quad & E[|\sum_{i=1}^{n}(r_{ij}-\mu_i)x_i|] \\
st \quad & \\
& \sum_{i=1}^{n}\mu_i x_i = \rho \\
& \sum_{i=1}^{n} x_i = 1 \\
& x_i = 0 \text{ or } \ell_i \le x_i \le u_i, \quad i=1,\ldots,n \\
& |supp(x)| \le K,
\end{aligned} \quad (7)$$

If we use $n$ additional binary variables we can re-write the LAMAD model as the following MILP:

$$\begin{aligned}
\min \quad & \tfrac{1}{T}\sum_{j=1}^{T} d_j \\
st \quad & \\
& d_j \ge \sum_{i=1}^{n}(r_{ij}-\mu_i)x_i \\
& -d_j \le \sum_{i=1}^{n}(r_{ij}-\mu_i)x_i \quad j=1,\ldots,T \\
& \sum_{i=1}^{n}\mu_i x_i = \rho \\
& \sum_{i=1}^{n} x_i = 1 \\
& \sum_{i=1}^{n} y_i \le K \\
& \ell_i y_i \le x_i \le u_i y_i && i=1,\ldots,n \\
& x_i \ge 0 && i=1,\ldots,n \\
& y_i \in \{0,1\} && i=1,\ldots,n \\
& d_j \ge 0 && j=1,\ldots,T
\end{aligned} \quad (8)$$

This problem has $n+T$ continuous variables, $n$ binary variables and $n+2T+3$ constraints. As before, the lower ($\ell_i$) and upper ($u_i$) bounds are set to 0.01 and 1, respectively. The frontiers have been computed for several (equally spaced) values of $\rho$ in $[\rho_{min},\rho_{max}]$, where the interval is determined as described in the previous subsection. In Fig. 5 we report the optimal solutions in the MAD-return plane, computed using CPLEX, for some instances based on real-world data and for some values of $K$.

### 2.2.3 Literature on Mixed Integer LP portfolio models

The need of solving large portfolio problems with real-world constraints justifies a long tradition in the literature of mixed integer LP portfolio models. Konno [29] tackles the portfolio optimization model, using a piecewise linear risk function. Historical monthly data of 50 stocks for 5 years are used. Konno and Yamazaki [32] compare the MAD model ex-post performance with that obtained by the MV model and the Single Index Model using 224 stocks with monthly data for 5 years. Linear programming based heuristics are used by Speranza [54], considering the negative semiMAD model with cardinality constraints, transaction costs and minimum transaction units. The minimax model has been presented by Young [59], who also describes how this model can be adapted to include linear transaction costs. A set of historical data of 7 stock indices is examined using the minimax and mean-variance portfolio selection rules. A comparison between these model is done, evaluating the optimal solution on 30 monthly data and testing the performance on the following 30 months. Bertsimas *et al.* [8] implement a mixed-integer programming model, using CPLEX as a solver, to construct a portfolio that is close to a target portfolio and controls frictional costs with cardinality constraints. Mansini and Speranza [37] show that finding a feasible solution to the portfolio selection problem using the Mean Semi-absolute Deviation model with roundlots is NP-complete. Some computational experiences are performed on data sets with at most 277 securities on a time period of 2 years. Different mixed integer linear programming models are presented by Kellerer *et al.* [27]. They compare the solutions of the semi-absolute deviation with fixed costs and possibly minimum lots, obtained by heuristic procedures and by CPLEX. The computational results are performed on the monthly rates of return of 244 securities for a time interval of 3 years. Konno and Wijayanayake [31] suggest a branch and bound algorithm for calculating a globally optimal solution of a portfolio problem, using MAD as risk measure under concave transaction costs and minimal transaction unit constraints. Some numerical tests are done with at most 60 monthly data of 200 stocks. Chiodi *et al.* [15] present a mixed integer linear programming model to solve a portfolio selection problem on mutual funds. Some heuristics are proposed, testing the problems on the historical data consisting of 310 mutual funds over the time period of 36 monthly returns for each fund. A portfolio selection problem with transaction costs and integer constraints on the quantities selected for the securities is presented by Mansini and Speranza [38].

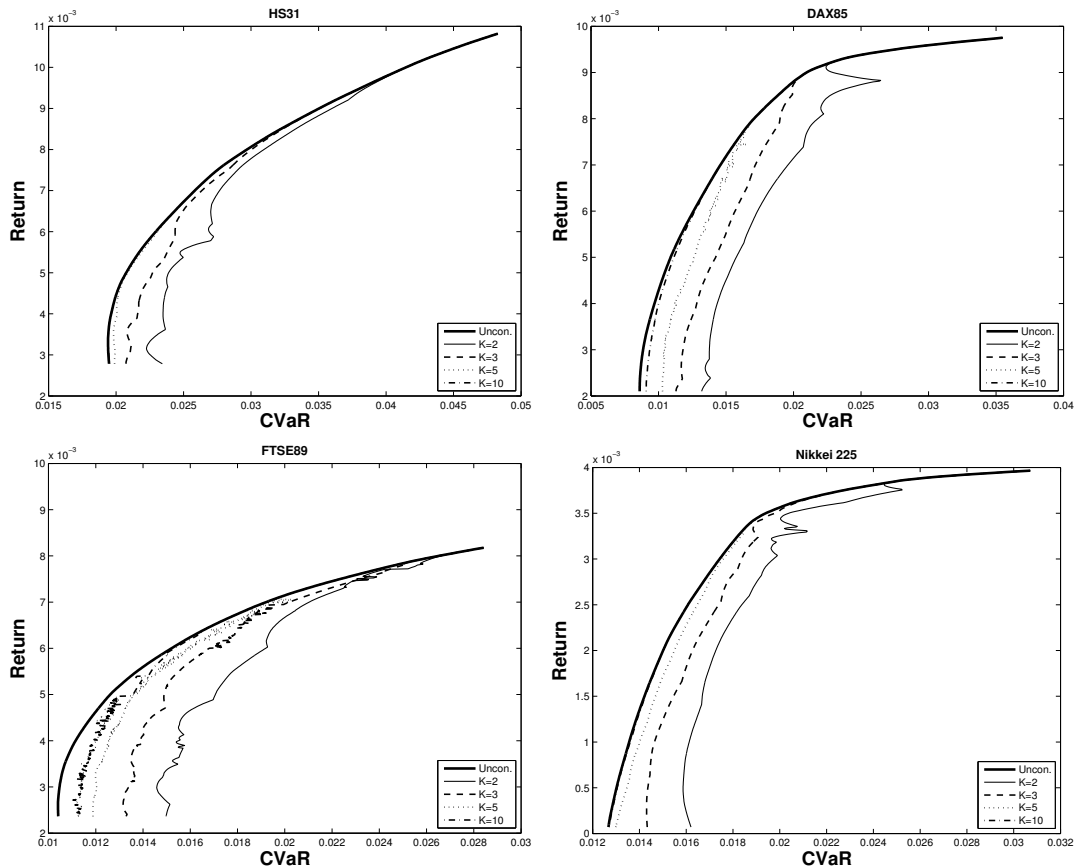

Figure 5: Frontiers of the *Limited Asset* MAD model

The number of stocks considered in the experiments are between 50 and 1000 on a time period varying from 2 to 6 years (this means that $100 \leq T \leq 300$).

While the MAD model with integer constraints has been widely analyzed in the literature, to the best of our knowledge few papers has been devoted to the CVaR model with real-world constraints. A financial and computational comparison of MAD and CVaR models with real features has been performed by Angelelli *et al.* [3]. Two different mixed integer linear programming models with integer stock units, transaction costs and a cardinality constraint are taken into account, analyzing their performance on real size instances. They use 104 weekly rates of return for the in-sample analysis and they test the performance of the optimal solutions obtained on the following 52 weeks.

## 3 Reduction to a Standard Quadratic Programming Problem

We propose here a new method for solving (2) that avoids the explicit use of additional binary variables. Our approach is based on the reduction of the LAM model (2) to a Standard Quadratic Programming (StQP) problem, as defined by Bomze [11], and is able to solve to optimality Beasley's problems and problems of greater dimension.

A StQP is the problem of minimizing a (possibly indefinite) quadratic form over the standard simplex $\Delta$, that is

$$\begin{aligned} \text{Min} \quad & x'Qx \\ \text{st} \quad & \\ & x \in \Delta = \{x \in \mathbb{R}^n : \sum_{i=1}^{n} x_i = 1, x_i \geq 0, \quad i = 1, \ldots, n\} \end{aligned} \quad (9)$$

Despite its formal simplicity, this problem is theoretically difficult to solve (NP-hard) when $Q$ is indefinite [11]. Indeed, its actual optimal solution for instances with more than 40 variables have not been reported in the literature until the recent paper by [51], where instances with more than 1000 variables have been solved.

We also point out that there is no loss of generality in restricting to quadratic *forms* instead of considering a general quadratic objective function. Indeed, over $\Delta$ a quadratic function $f(x) = x'Px + 2q'x$ coincides with the quadratic form $x'Qx$, where $Q = P + eq' + qe'$, and $e$ denotes the all-ones vector.

Problem (1) can be easily transformed into a (convex) StQP problem by using a quadratic penalty for the return constraint:

$$\begin{aligned} \text{Min} \quad & f_M(x) = \sum_{i=1}^{n} \sum_{j=1}^{n} \sigma_{ij} x_i x_j + M[\sum_{i=1}^{n} \mu_i x_i - \rho]^2 \\ \text{st} \quad & \\ & x \in \Delta \end{aligned} \quad (10)$$

Adding the cardinality constraint $|supp(x)| \leq K$ to (10) amounts to minimizing $f_M$ on the faces of dimension not greater than $K$ of the standard simplex $\Delta$. If we further add the condition that $\ell_i \leq x_i \leq u_i$ for $i = 1, \ldots, n$, we obtain a StQP with cardinality and upper and lower bound constraints which is equivalent to (4).

In the next section we describe how to solve a StQP with cardinality and upper and lower bound constraints by adapting the algorithms developed by [51] for the unconstrained case.

## 4 Theoretical Results and Solution Method

We consider the cardinality constrained StQP problem:

$$\begin{aligned}
\min \quad & f(x) = x'Qx \\
\text{st} \quad & \\
& x \in \Delta = \{x \in \mathbb{R}^n : \sum_{i=1}^n x_i = 1, x_i \geq 0, \quad i = 1, \ldots, n\} \\
& |supp(x)| \leq K
\end{aligned} \quad (11)$$

In order to restrict the search for its global minimizers, we use the following QP extension of the fundamental theorem of Linear Programming.

**Theorem 3** *[51, 57, 58]. A quadratic function $f$ that is bounded below on a (pointed) polyhedron $P$ attains its minimum on $P$ in the relative interior of a face of $P$ where $f$ is strictly convex.*

Let $N = \{1, \ldots, n\}$. Every face of $\Delta$ has the form $\Delta_I = \{x \in \Delta : \sum_{i \in I} x_i = 1\}$, where $I \subseteq N$ is a subset of indices. Furthermore, the dimension $\dim(\Delta_I)$ of $\Delta_I$ coincides with the cardinality $|I|$ of $I$. Let $\mathcal{I}_K$ denote the family of all subsets of $N$ with cardinality at most $K$. Then the cardinality constrained StQP (11) can be reformulated as:

$$\min_{x \in \bigcup_{I \in \mathcal{I}_K} \Delta_I} f(x) = x'Qx \quad (12)$$

Hence we obtain the following straightforward consequence of Theorem 3:

**Corollary 4** *At least one global minimizer of (11) must be in the relative interior of a face $\Delta_I$ of $\Delta$ where $f$ is strictly convex and $|I| \leq K$.*

For $I \subseteq N$, let $Q_I$ denote the submatrix of $Q$ formed by those elements with row and column indices in $I$. When $Q_I$ is positive definite, the unique global minimizer of the quadratic form $x'Q_I x$ on the hyperplane $\sum_{i \in I} x_i = 1$ is attained at the point $x_I^* = (e'Q_I^{-1}e)^{-1}Q_I^{-1}e$. Thus the quadratic form $f(x)$ has a global minimizer on $\Delta$ in the relative interior $rint(\Delta_I)$ of a face $\Delta_I$ where $f$ is strictly convex only if $x_I^* \in rint(\Delta_I)$.

To every subset $I \subseteq N$ we associate the (nonlinear) weight

$$w(I) = \min\{f(x) : x \in \Delta_I\}.$$

Corollary 4 and simple matrix algebra imply that

$$\min_{x \in \bigcup_{I \in \mathcal{I}_K} \Delta_I} x'Qx = \min_{I \in \mathcal{C}_K} w(I) = \min_{I \in \mathcal{C}_K} f(x_I^*) = \min_{I \in \mathcal{C}_K} (e'Q_I^{-1}e)^{-1}, \quad (13)$$

where $\mathcal{C}_K$ is the subset of $\mathcal{I}_K$ defined by

$$\mathcal{C}_K = \{I \in \mathcal{I}_K : Q_I \text{ is positive definite and } x_I^* \in rint(\Delta_I)\}.$$

In view of (13), the cardinality constrained StQP could be solved by evaluating $(e'Q_I^{-1}e)^{-1}$ for all elements $I \in \mathcal{C}_K$, but this is clearly not practical for large values of $n$ and $K$. However, another recent theoretical result can be used to restrict the search for a global minimizer:

**Theorem 5** *[51]. If $x^*$ is a global minimizer of a quadratic function $f$ on a polyhedron $P$, then there exists a nested sequence of faces $F^1 \subset F^2 \subset \ldots \subset F^k$ of $P$, with dimension $dim(F^i) = i$, where $f$ is strictly convex, has an interior global minimizer $\hat{x}_{F^i}$, and $x^* = \hat{x}_{F^k}$.*

For $j \leq K$, let $\overline{\mathcal{C}}_j = \{I \in \mathcal{C}_K : |I| = j\}$. Then the above theorem guarantees that for any $I^*$ minimizing $w(I)$ on $\mathcal{C}_K$ there exists a sequence $I_1 \subseteq I_2 \subseteq \cdots \subseteq I_h = I^*$, such that $I_j \in \overline{\mathcal{C}}_j$ for all $j = 1, \ldots, h$. Thus we can apply the following algorithm to solve the cardinality constrained StQP (11) or, equivalently, to minimize $w(I)$ on $\mathcal{C}_K$:

## Increasing Set Algorithm

1 Set $\mathcal{C}_0 = \emptyset$, $\mathcal{C}_1 = \{\{i\}, i \in N\}$

2 MIN(1) $= \min_{I \in \mathcal{C}_1} w(I) = \min_{1 \leq i \leq n} q_{ii}$

3 **for** $j = 1$ to $K$

4     **do** construct $\overline{\mathcal{C}}_{j+1}$ by increasing, if possible, all elements in $\overline{\mathcal{C}}_j$

5         **if** $\overline{\mathcal{C}}_{j+1} = \emptyset$

6             **then** MIN($h$) = MIN($j$) for $h = j+1, \ldots, K$, **return**(MIN($K$))

7             **else** MIN($j+1$) = $\min\{\text{MIN}(j), \min_{I \in \overline{\mathcal{C}}_{j+1}} w(I)$

8                 **return** (MIN($K$))

Note that, by Theorem 5, at any iteration $j$, MIN($j$) contains the minimum value of $w(I)$ among all sets in $\mathcal{C}_j$. Furthermore, if $\overline{\mathcal{C}}_{j+1} = \emptyset$ in step 5, then, again by Theorem 5, $\overline{\mathcal{C}}_h$ must be empty for all $h \geq j+1$. Hence the algorithm correctly stops with the global minimizer in $\mathcal{C}_K$. In fact, at each iteration $j$ the Increasing Set Algorithm provides in MIN($j$) the solution to the StQP problem with cardinality constraint $|supp(x)| \leq j$.

We have proved that the Increasing Set Algorithm is exact. Unfortunately, it has exponential complexity in the worst case, and may be too slow in practice for large size problems. However, we obtain a very good heuristic if we bound at each iteration the size of $\overline{\mathcal{C}}_j$ by keeping only a limited number of sets $I$ with the best values of $w(I)$. From a theoretical viewpoint, we can achieve polynomial time complexity in this way, but of course we lose the guarantee of optimality. In practice, however, we have observed considerable reduction in the running time without losing optimality in all real-world instances, described in Section 5, that have been solved with both our algorithm and with CPLEX.

We should point out that in order to apply our algorithm to the (reformulated) LAM model that also includes lower and upper bounds $\ell_i$ and $u_i$ on the variables $x_i > 0$, we need to further modify the basic Increasing Set Algorithm described above. Indeed, to solve a StQP with cardinality and lower and upper bound constraints, we find the sets $\overline{\mathcal{C}}'_j$ and $\overline{\mathcal{C}}''_j$, where $\overline{\mathcal{C}}'_j = \{I \in \overline{\mathcal{C}}_j : \ell \leq x^*_I \leq u\}$ and $\overline{\mathcal{C}}''_j = \overline{\mathcal{C}}_j \setminus \overline{\mathcal{C}}'_j$. We replace $\min_{I \in \overline{\mathcal{C}}_{j+1}} w(I)$ in step 7 of the algorithm with $\min_{I \in \overline{\mathcal{C}}'_{j+1}} w(I)$, and we memorize the list of all sets $I$ in $\mathcal{C}''_K = \bigcup_{j=1}^K \overline{\mathcal{C}}''_j$. At the end of the algorithm, we then replace $\text{MIN}(K)$ with $\min\{\text{MIN}(K), \min_{I \in \mathcal{C}''_K} w(I)\}$. This can be done efficiently by observing that, for all $I \in \mathcal{C}''_K$, $w(I)$ can be computed by solving a convex quadratic programming problem of dimension $|I|$, and that we only need to solve such problems for those $I \in \mathcal{C}''_K$ for which $f(x^*_I) < \text{MIN}(K)$.

Furthermore, if we want to find the best portfolio among those that contain a given subset $J$ of assets (i.e., satisfy the *pre-assignment constraints* of Di Gaspero et al. [18]), then we just need to modify the Increasing Set Algorithm so that it starts with the family $\mathcal{C}_{|J|} = \{J\}$. Thus the pre-assignment constraints actually simplify the solution of the problem with the Increasing Set Algorithm, which needs less iterations to terminate.

## 5 Data Sets and Computational Results

### 5.1 Data Sets

An important issue when evaluating computational results for a class of problems is the availability of benchmark data sets, possibly with solutions, that can be used by researchers to compare the efficiency of their algorithms, and the quality of the solutions obtained in the case of heuristics. Unfortunately, in the case of the LAM model, such benchmark data sets are currently only partially available.

The most popular publicly available data sets based on real-world data for the LAM model seem to be the ones described by Chang et al. [14]. They include covariance matrices and expected return vectors of sizes ranging from 31 to 225 built from weakly price data from March 1992 to September 1997 for the Hang Seng, DAX, FTSE 100, S&P 100, and Nikkei 225 capital market indices. The weakly price data are contained in the files indtrack1, indtrack2,..., indtrack5 available from Beasley's OR-Library [7] at http://people.brunel.ac.uk/~mastjjb/jeb/orlib/indtrackinfo.html, where one can also find weakly price data for the S&P 500 (457 assets), Russell 2000 (1318 assets), and Russell 3000 (2151 assets) capital market indices in the files indtrack6, indtrack7, indtrack8, as described in [13]. The return rates for these eight markets have been computed as logarithmic variations of the quotation prices $(\ln(P_t/P_{t-1}))$. The historical realizations consist of 290 rates of return. It should be mentioned however that, for commercial reasons, the data sets have been anonymized, in the sense that the names of the stocks associated to the data are not disclosed. Thus we decided to construct, and to make available in the web page http://w3.uniroma1.it/Tardella/datasets.html, five additional data sets that refer to the EuroStoxx50 (Europe), FTSE 100 (UK), MIBTEL (Italy), S&P 500 and NASDAQ (USA) capital market indices. These data sets contain the names of all the stocks included. For each stock we obtained 263 weakly price data, adjusted for dividends, from Yahoo Finance for the period from March 2003 to March 2008. Stocks with more than two consecutive missing values were disregarded. The missing values of the remaining stocks were interpolated. We thus obtained data sets of 47 stocks for EuroStoxx50, 76 for FTSE 100, 221 for MIBTEL, of 476 for S&P 500, and 2191 for NASDAQ. We then computed (logarithmic) weekly returns, expected returns, and covariance matrices based on the (in-sample) data for the period March 2003 - March 2007. The remaining data, for the period April 2007 - March 2008, have been used as out-of-sample data to evaluate the ex-post performance of the portfolios obtained with the models (see Section 6).

|  |  | Number of | $K = 5$ | | $K = 10$ | |
|---|---|---|---|---|---|---|
|  |  | assets ($N$) | CPLEX | INCR. SET | CPLEX | INCR. SET |
| OR-Library | Hang Seng | 31 | **8** | 37 | **6** | 55 |
|  | DAX 100 | 85 | 1015 | **377** | **136** | 797 |
|  | FTSE 100 | 89 | 2978 | **663** | **986** | 1750 |
|  | S&P 100 | 98 | 197816 | **1223** | 85912 | **4184** |
|  | Nikkei | 225 | **161** | 375 | **61** | 752 |
|  | S&P 500 | 457 | - | **5863** | - | **19710** |
|  | Russell 2000 | 1318 | - | **12172** | - | **14611** |
|  | Russell 3000 | 2151 | - | **48098** | - | **52643** |
|  | EuroStoxx50 | 47 | **30** | 165 | **17** | 341 |
|  | FTSE 100 | 76 | 354 | **345** | **79** | 717 |
|  | MIBTEL | 221 | - | **3603** | - | **19220** |
|  | S&P 500 | 476 | - | **7510** | - | **45491** |
|  | NASDAQ | 2191 | - | **55500** | - | **63671** |

Table 1: Running times in seconds to solve the LAM model for 500 return values with $K$ assets

A drawback of Beasley's data sets is the lack of optimal (or best known) solutions to the LAM model based on them, although some statistics and some indicators that measure the quality of the solutions obtained are presented in [14, 17, 18, 26, 45, 50]. We fill this gap by providing the optimal (or best known) solutions to the LAM model both for our data sets and for the ones contained in Beasley's OR-Library.

## 5.2 Computational Results for the LAM Model

In this section we provide some computational results comparing our heuristic algorithm with the exact MIQP solver in CPLEX 11.0. We point out that although optimality is not guaranteed for our algorithm, we have observed that in all instances where CPLEX could solve the problem, the solutions found by the two algorithms coincided up to numerical precision. Hence we need not report the accuracy of the solutions found by our algorithm.

Our algorithm is coded in MATLAB 7.4 and executed on a workstation with Intel Core2 Duo CPU (T7500, 2.2 GHz, 4Gb RAM) under Windows Vista. CPLEX 11.0 is also called from MATLAB with the TOMLAB/CPLEX toolbox [25].

For each data set, we computed $\rho_{min}$ and $\rho_{max}$ as described in Section 2.1 by solving the classical (unconstrained) Markowitz model. We then repeatedly solved the LAM model (4) for 500 equally spaced returns between $\rho_{min}$ and $\rho_{max}$ thus obtaining 500 values of the function $\phi_K(\rho)$. A simple post-processing of these values allowed us to compute $\phi'_K(\rho)$, and to determine the points on the Efficient Frontier of the LAM model (4), also called LAMEF. The graphs obtained for some data sets are shown in Fig. 6.

As in [14, 17, 18, 26, 39, 45, 50], we report results for problems with cardinality constraints $K = 5$ and $K = 10$, lower bound $\ell_i = 0.01$, and upper bound $u_i = 1$ for all $i = 1, \ldots, n$. The choice of $K = 10$ as the largest cardinality constraint is also justified by the observation that for several data sets the optimal portfolio in the classical Markowitz model does not include more than 10 stocks for more than half of the $\rho$ values (see also Fig. 7). Furthermore, we observed that the number of stocks with positive weight in the optimal portfolio for the classical Markowitz model might be an important indicator of the practical computational complexity for most exact algorithms for the LAM model. This is certainly the case both for CPLEX and for our Increasing Set algorithm, as clearly results by comparing in Table 1 the computation time for S&P 100, (98 assets) with the one for Nikkei (225 assets). The computation for S&P 100 takes much longer because it has many more assets in the optimal portfolio for the classical Markowitz model, as shown in Fig. 7. In fact, for Nikkei the number of assets in the optimal portfolio is less than or equal to 10 for about half of the target return values, so that the cardinality constraint is not active. The maximum number of assets in the Markowitz portfolios is 15. For the S&P100 data set the maximum number of assets in a portfolio is 34 and only for about 33% of the target return values the Markowitz portfolio contains less than 10 assets. Indeed, when the cardinality constraint is not active, the *unconstrained* and the *Limited Asset* Markowitz Efficient Frontier (LAMEF) coincide, and both CPLEX and the Increasing Set Algorithm have no difficulties to solve the LAM model. Hence, the hardness of computation of the LAMEF seems to be related not only to the number of variables but also to the number of assets satisfying $x_i > 0$ in the solution of the unconstrained Markowitz model.

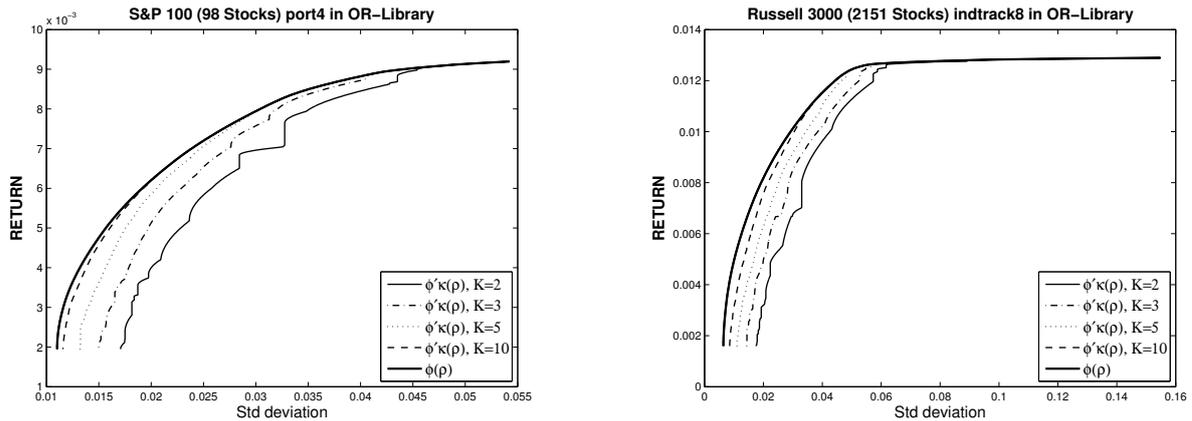

Figure 6: Examples of Efficient Frontiers of the LAM model

We should make some remarks concerning the running times presented in Table 1. First, the Increasing Set algorithm is currently a prototype algorithm coded in MATLAB tailored for the LAM model, while the solver in CPLEX is a highly optimized general purpose MIQP solver. Furthermore, the times reported to solve the LAM model for a given $K$ with the Increasing Set algorithm should actually be read as the times required to solve the model for all $K' \leq K$, as observed in Section 4. Thus such times are clearly increasing with $K$, but they refer to solving a family of problems. On the other hand, the running times of CPLEX seem to almost always decrease with $K$. Hence, CPLEX might be used as a complementary tool with respect to the Increasing Set algorithm. However, it should be noted that CPLEX is currently unable to solve the largest problems in our data sets.

Some authors [17, 18, 45, 50] have measured the quality of the results obtained by their heuristic algorithms by computing an *Average Percentage Loss* (APL) comparing the risk obtained by the algorithms for the LAM model with a given required return $\rho$ to the optimal risk for the same return in the classical (unconstrained) Markowitz model. Since the definitions of APL considered by these authors are slightly different, we compare our results separately. More precisely, with our notation, the APL considered by Moral-Escudero *et al.* [45] is defined as

$$\text{APL}_1 = \sum_{j=1}^{100} \frac{\phi_K(\rho_j) - \phi(\rho_j)}{\phi(\rho_j)}, \tag{14}$$

where $\phi_K(\rho_j)$ is the optimal value function of problem (2), the returns $\rho_j$, with $j = 1,\ldots,100$, are equally distributed in the interval $[\rho_{min}, \rho_{max}]$, $K = 10$, $\ell_i = 0.01$ and $u_i = 1$ for all $i$. While, the APL considered in [17, 18] and in [50] is obtained as

$$\text{APL}_2 = \sum_{j=1}^{100} \frac{\phi'_K(\rho_j) - \phi(\rho_j)}{\phi(\rho_j)}, \tag{15}$$

where $\phi'_K(\rho_j)$ is the optimal value function of problem (2), where the equality constraint on the target expected return is replaced by $\sum_{i=1}^{n} \mu_i x_i \geq \rho$.

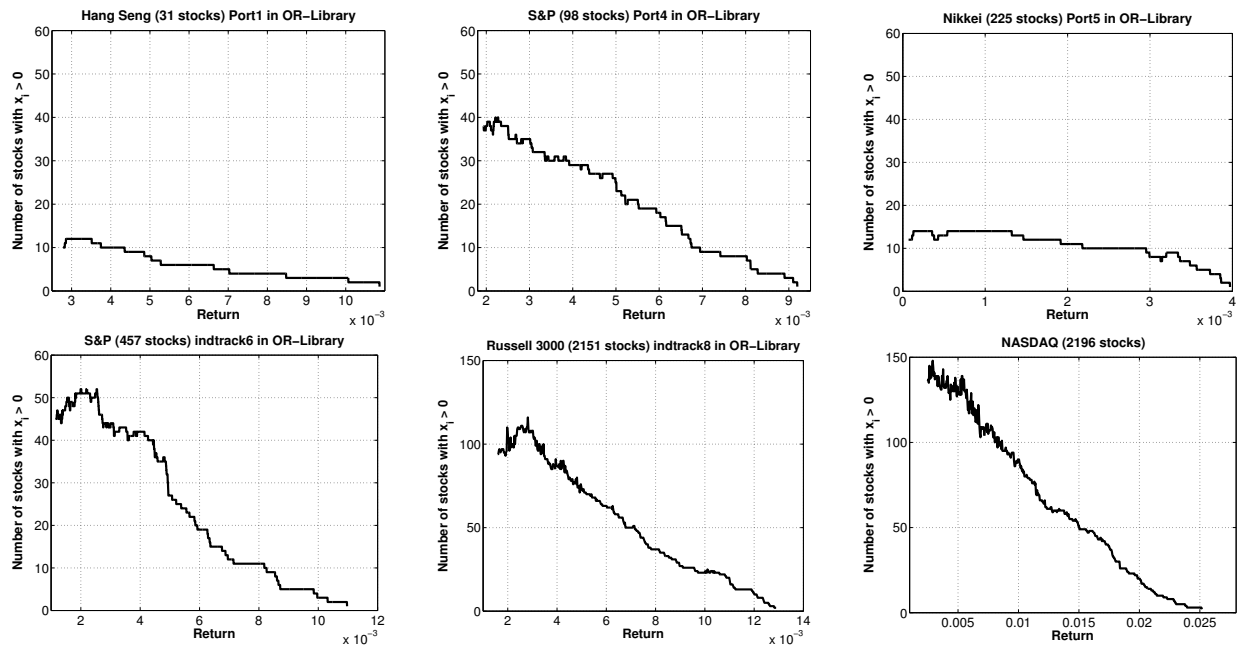

Figure 7: Number of assets in the unconstrained Mean-Variance optimal portfolio

Since we could compute the exact values of $\phi_K(\rho_j)$ and $\phi'_K(\rho_j)$, the Average Percentage Loss that we obtain is the best possible with respect to the return values used (we call it Exact APL). The results that we obtain show that, in spite of the theoretical difference, the values of $APL_1$ and $APL_2$ are the same or almost the same for all the data sets that we consider. Table 2 shows a comparison between the results of the exact $APL_1$ computed by the Increasing Set Algorithm with the APL reported in [45]. Similarly, we compare in Table 3 the results of the exact $APL_2$, with the values computed by Di Gaspero *et al.* [18] and by Schaerf [50]. We remark that the values that we obtain are slightly better than those reported by the other authors. This seems to contrast with the optimality of the results claimed by Di Gaspero *et al.* [18]. However, the small differences in the results might also be explained by the choice of the points on the frontier, or by the numerical precision of the algorithms. In order to make the comparison with our results easier, we have made available in the web page http://w3.uniroma1.it/Tardella/APL.html the 100 return values and the covariance matrices obtained from Beasley's OR Library that we used in our computations of the APL, together with the optimal solutions $\phi_K(\rho)$ and $\phi'_K(\rho)$ that we found for problem (2) with equality or inequality constraint on the expected return.

| Data set | # of Assets | Exact $APL_1$ | [45] |
|---|---|---|---|
| Hang Seng | 31 | 0.00312 | 0.00321 |
| DAX 100 | 85 | 2.50749 | 2.53180 |
| FTSE 100 | 89 | 1.90225 | 1.92150 |
| S&P 100 | 98 | 4.64937 | 4.69507 |
| Nikkei | 225 | 0.19978 | 0.20198 |

Table 2: Comparison of Average Percentage Loss $APL_1$

| Data set | # of Assets | Exact $APL_2$ | [18] | [50] |
|---|---|---|---|---|
| Hang Seng | 31 | 0.00312 | 0.00321 | 0.00409 |
| DAX 100 | 85 | 2.50742 | 2.53139 | 2.53617 |
| FTSE 100 | 89 | 1.90203 | 1.92146 | 1.92597 |
| S&P 100 | 98 | 4.64937 | 4.69371 | 4.69816 |
| Nikkei | 225 | 0.19978 | 0.20199 | 0.20258 |

Table 3: Comparison of Average Percentage Loss $APL_2$

### 5.3 Computational Results for the LACVaR and LAMAD Models

In this section we report the running times required by the commercial solver CPLEX to find exact or approximate (within a specified tolerance) solutions to the LACVaR and LAMAD models for the instances considered in Section 5.2 for the LAM model with the same settings:

- the lower bound $\ell_i$ and upper bound $u_i$ are set to 0.01 and 1, respectively;

- the maximum number $K$ of securities in the portfolio (cardinality constraint) is fixed at 5 and 10;

- for each data set we vary $\rho$ in the interval $[\rho_{min}, \rho_{max}]$ defined in Section 2.1 for the LAM model.

We have solved the LACVaR (6) and LAMAD (8) models for 500 equally spaced returns between $\rho_{min}$ and $\rho_{max}$, obtaining 500 points for each frontier. The graphs for some data sets are shown in Figs. 4 (LACVaR) and 5 (LAMAD).

In Tables 4 and 5, for each data set we report the number of assets, variables, and constraints, and the corresponding running times. We point out that the number of variables and constraints in the LACVaR and LAMAD models depends not only on the number of assets $n$ but also on the length of the in-sample period $T$ chosen ($T = 290$ for the OR-Library data sets and $T = 210$ for our data sets). Thus, the number of variables tends to become fairly large making the mixed integer problem difficult to solve. In Tables 4 and 5 we present the running times of CPLEX for finding both optimal (denoted by OPT) and approximate (denoted by APPR) solutions. The optimal values are obtained using default tolerances ($10^{-6}$), while the approximate values are computed relaxing to $10^{-4}$ the absolute tolerance on the gap found by CPLEX. We also report the running times for finding the optimal solutions to the unconstrained models (Uncon).

|  | **LACVaR** | Number of | | | | $K = 5$ | | $K = 10$ | |
|---|---|---|---|---|---|---|---|---|---|
|  |  | assets (N) | variables | constraints | Uncon | OPT | APPR | OPT | APPR |
| OR-Library | Hang Seng | 31 | 353 | 324 | 15 | 20 | 19 | 19 | 18 |
|  | DAX 100 | 85 | 461 | 378 | 22 | 670 | 530 | 155 | 45 |
|  | FTSE 100 | 89 | 469 | 382 | 22 | 9097 | 6440 | 3753 | 216 |
|  | S&P 100 | 98 | 487 | 391 | 24 | - | - | - | 7904 |
|  | Nikkei | 225 | 613 | 454 | 33 | 934 | 394 | 198 | 101 |
|  | S&P 500 | 457 | 1205 | 750 | 97 | - | - | - | - |
|  | Russell 2000 | 1318 | 2927 | 1611 | 577 | - | - | - | - |
|  | Russell 3000 | 2151 | 4593 | 2444 | 648 | - | - | - | - |
|  | EuroStoxx50 | 47 | 305 | 260 | 14 | 77 | 62 | 30 | 21 |
|  | FTSE 100 | 76 | 363 | 289 | 17 | 337 | 238 | 65 | 38 |
|  | MIBTEL | 221 | 653 | 434 | 26 | 13773 | 9393 | 21666 | 1061 |
|  | S&P 500 | 476 | 1163 | 689 | 42 | - | - | - | - |
|  | NASDAQ | 2191 | 4593 | 2404 | 470 | - | - | - | - |

Table 4: Running times in seconds to solve the LACVaR model for 500 return values with $K$ assets

|  | **LAMAD** | Number of | | | | $K = 5$ | | $K = 10$ | |
|---|---|---|---|---|---|---|---|---|---|
|  |  | assets (N) | variables | constraints | Uncon | OPT | APPR | OPT | APPR |
| OR-Library | Hang Seng | 31 | 352 | 614 | 22 | 136 | 90 | 95 | 68 |
|  | DAX 100 | 85 | 460 | 668 | 56 | 4956 | 4856 | 13173 | 488 |
|  | FTSE 100 | 89 | 468 | 672 | 64 | - | 16255 | - | 288 |
|  | S&P 100 | 98 | 486 | 681 | 67 | - | 6057 | - | 723 |
|  | Nikkei | 225 | 740 | 744 | 66 | 1951 | 661 | 814 | 553 |
|  | S&P 500 | 457 | 1204 | 1040 | 235 | - | 4823 | - | 1407 |
|  | Russell 2000 | 1318 | 2926 | 1901 | 1148 | - | - | - | - |
|  | Russell 3000 | 2151 | 4592 | 2734 | 2090 | - | - | - | - |
|  | EuroStoxx50 | 47 | 304 | 470 | 24 | 341 | 124 | 112 | 79 |
|  | FTSE 100 | 76 | 362 | 499 | 32 | 11299 | 1337 | 14980 | 178 |
|  | MIBTEL | 221 | 652 | 644 | 71 | - | 9809 | - | 1536 |
|  | S&P 500 | 476 | 1162 | 899 | 130 | - | 2103 | - | 900 |
|  | NASDAQ | 2191 | 4592 | 2614 | 1148 | - | - | - | - |

Table 5: Running times in seconds to solve the LAMAD model for 500 return values with $K$ assets

In Table 6 we summarize the running times of the Increasing Set (IS) algorithm for the LAM model and of CPLEX, both with small tolerance (OPT) and with larger tolerance (APPR), for all models. The values are missing when the code has not been able to solve the problem within 2 days. From the table it appears that, apart from the simplest problems, the running time of the Increasing Set algorithm is comparable to that of CPLEX with large tolerance (APPR). However, CPLEX is not able to solve, even approximatively, the largest problems. Furthermore, it should be recalled that the Increasing Set algorithm actually finds a solution for all $K' \leq K$, while CPLEX solves the problem for a single $K$. Thus from the computational viewpoint it turns out that quite surprisingly in most cases the quadratic LAM model can be solved more efficiently with our algorithm than the linear LAMAD and LACVaR models with current state-of-the-art solvers. Therefore, unless more efficient ad-hoc algorithms are developed for the mixed integer linear models, the LAM model should be preferred for large size problems.

|  |  | N of | LAM | | | | LAMAD | | | | LACVaR | | | |
|---|---|---|---|---|---|---|---|---|---|---|---|---|---|---|
|  |  |  | $K=5$ | | $K=10$ | | $K=5$ | | $K=10$ | | $K=5$ | | $K=10$ | |
|  |  | assets | CPLEX | IS | CPLEX | IS | OPT | APPR | OPT | APPR | OPT | APPR | OPT | APPR |
| OR-Library | Hang Seng | 31 | 8 | 37 | 6 | 55 | 136 | 90 | 95 | 68 | 20 | 19 | 19 | 18 |
|  | DAX 100 | 85 | 1015 | 377 | 136 | 797 | 4956 | 4856 | 13173 | 488 | 670 | 530 | 155 | 45 |
|  | FTSE 100 | 89 | 2978 | 663 | 986 | 1750 | - | 16255 | - | 288 | 9097 | 6440 | 3753 | 216 |
|  | S&P 100 | 98 | 197816 | 1223 | 85912 | 4184 | - | 6057 | - | 723 | - | - | - | 7904 |
|  | Nikkei | 225 | 161 | 375 | 61 | 752 | 1951 | 661 | 814 | 553 | 934 | 394 | 198 | 101 |
|  | S&P 500 | 457 | - | 5863 | - | 19710 | - | 4823 | - | 1407 | - | - | - | - |
|  | Russell 2000 | 1318 | - | 12172 | - | 14611 | - | - | - | - | - | - | - | - |
|  | Russell 3000 | 2151 | - | 48098 | - | 52643 | - | - | - | - | - | - | - | - |
|  | EuroStoxx50 | 47 | 30 | 165 | 17 | 341 | 341 | 124 | 112 | 79 | 77 | 62 | 30 | 21 |
|  | FTSE 100 | 76 | 354 | 345 | 79 | 717 | 11299 | 1337 | 14980 | 178 | 337 | 238 | 65 | 38 |
|  | MIBTEL | 221 | - | 3603 | - | 19220 | - | 9809 | - | 1536 | 13773 | 9393 | 21666 | 1061 |
|  | S&P 500 | 476 | - | 7510 | - | 45491 | - | 2103 | - | 900 | - | - | - | - |
|  | NASDAQ | 2191 | - | 55500 | - | 63671 | - | - | - | - | - | - | - | - |

Table 6: Running times in seconds to solve the Limited Asset models for 500 return values with $K$ assets

## 6  Evaluation of Ex-post Performance

Out-of-sample experiments allow evaluation of the effectiveness of portfolio models for actual risk management purposes. The computed portfolios are built by solving the unconstrained models and the *Limited Asset* ones ($K = 5, 10$; $\ell_i = 0.01, u_i = 1$) on a given sample interval [March 2003, March 2007] for a fixed value of the target return $\rho$. After that, we simulate the holding of such portfolios for the time interval [April 2007, March 2008], and we evaluate their ex-post performance using out-of-sample data for the EuroStoxx 50, FTSE 100, MIBTEL, S&P500 and NASDAQ capital markets. Such performances are compared to that of the official capital market index in the same period. Some results are illustrated in Figs. 8 and 9. In Fig. 8 we show the results obtained for low risk strategies ($\rho = \rho_{min}$). These results seem to indicate that, although the portfolios selected by the *Limited Asset* models contain at most 5 or 10 securities, they have a better performance than the ones provided by the unconstrained models.

We also notice that the performance of the LAM and LAMAD portfolios seems to be at least as good as that of the LACVaR portfolio. Moreover, all the portfolios found by the models generally present a better performance than the market index in longer-term investment horizons.

In Fig. 9 we present similar experiments for higher risk strategies ($\rho = \rho_{min} + 1/2(\rho_{max} - \rho_{min})$). For this target return level, it seems that the assets limitation for the linear models is less important for the portfolio composition. Indeed, the performance of the unconstrained MAD and CVaR portfolios are similar to the constrained ones, while the optimal portfolios for the LAM model with high values of $\rho$ still provide portfolios with different performance. This is because the Markowitz model leads to a more diversified portfolio than the MAD and CVaR models [36].

## 7  Conclusions and Further Research

In this paper we have presented an efficient algorithm for a Mean-Variance portfolio selection model with constraints, coming from real-world practice, that are difficult to handle computationally. With this algorithm we can solve to optimality not only some well-known benchmark problems, but also larger problems with more than 2000 variables.

Our algorithm is based on a completely new approach that starts from a pair of assets and tries to add one asset at a time in an optimal manner by exploiting some recent theoretical results on Quadratic Programming.

We have also analyzed the CVaR and MAD models with cardinality constraints and buy-in thresholds. These are mixed integer linear programming (MILP) models, that have been solved using CPLEX, a state-of-the-art commercial solver. Although one expects a MILP model to be more tractable than a MIQP one, the computational results have shown that CPLEX requires more time to solve the *Limited Asset* CVaR and MAD models than the time needed by our algorithm to solve the *Limited Asset* Markowitz one, particularly

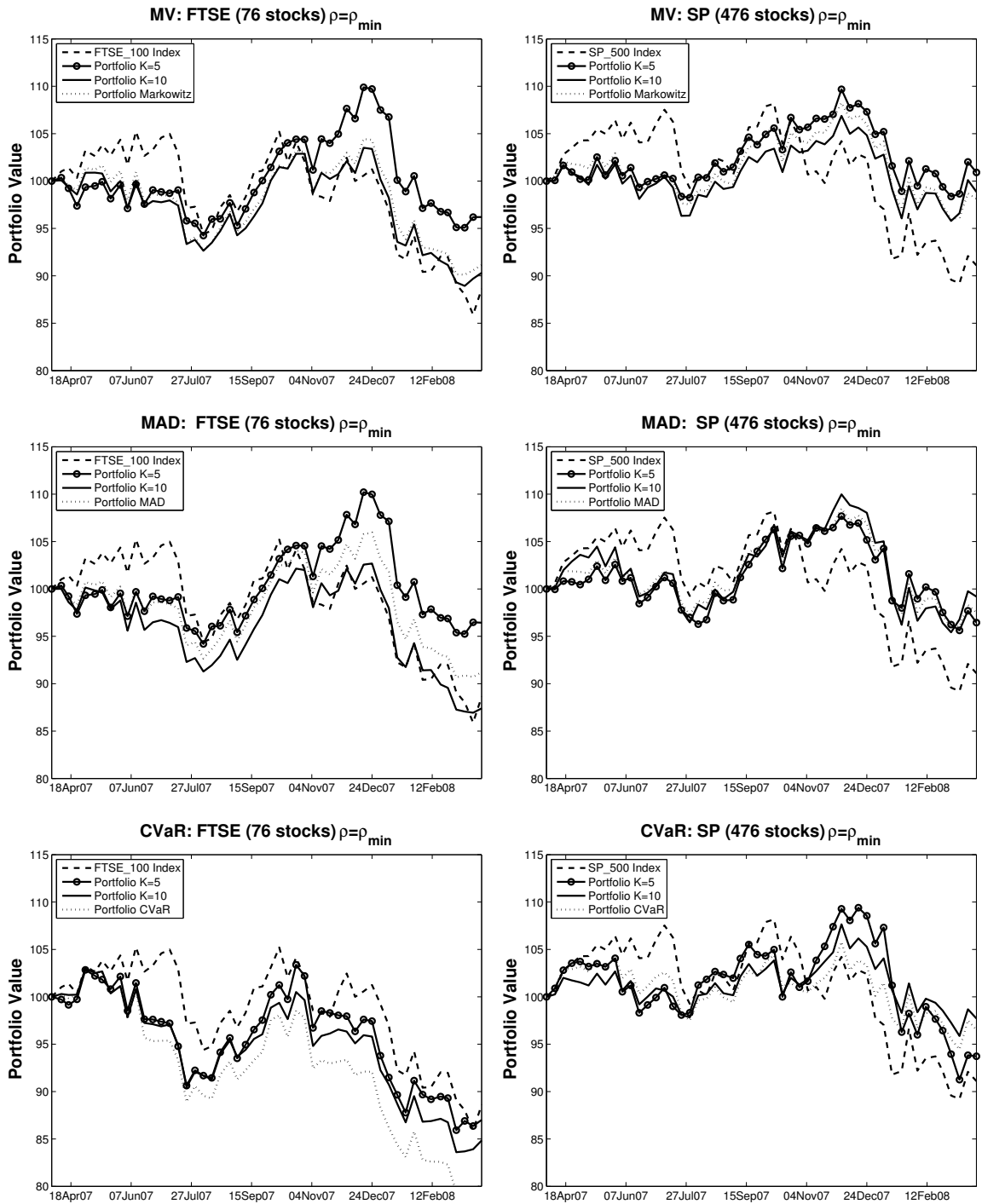

Figure 8: Evaluation of ex-post performances with $\rho = \rho_{min}$ for the Markowitz (top), MAD (middle) and CVaR (bottom) models on the FTSE100 (left) and SP500 (right) data sets

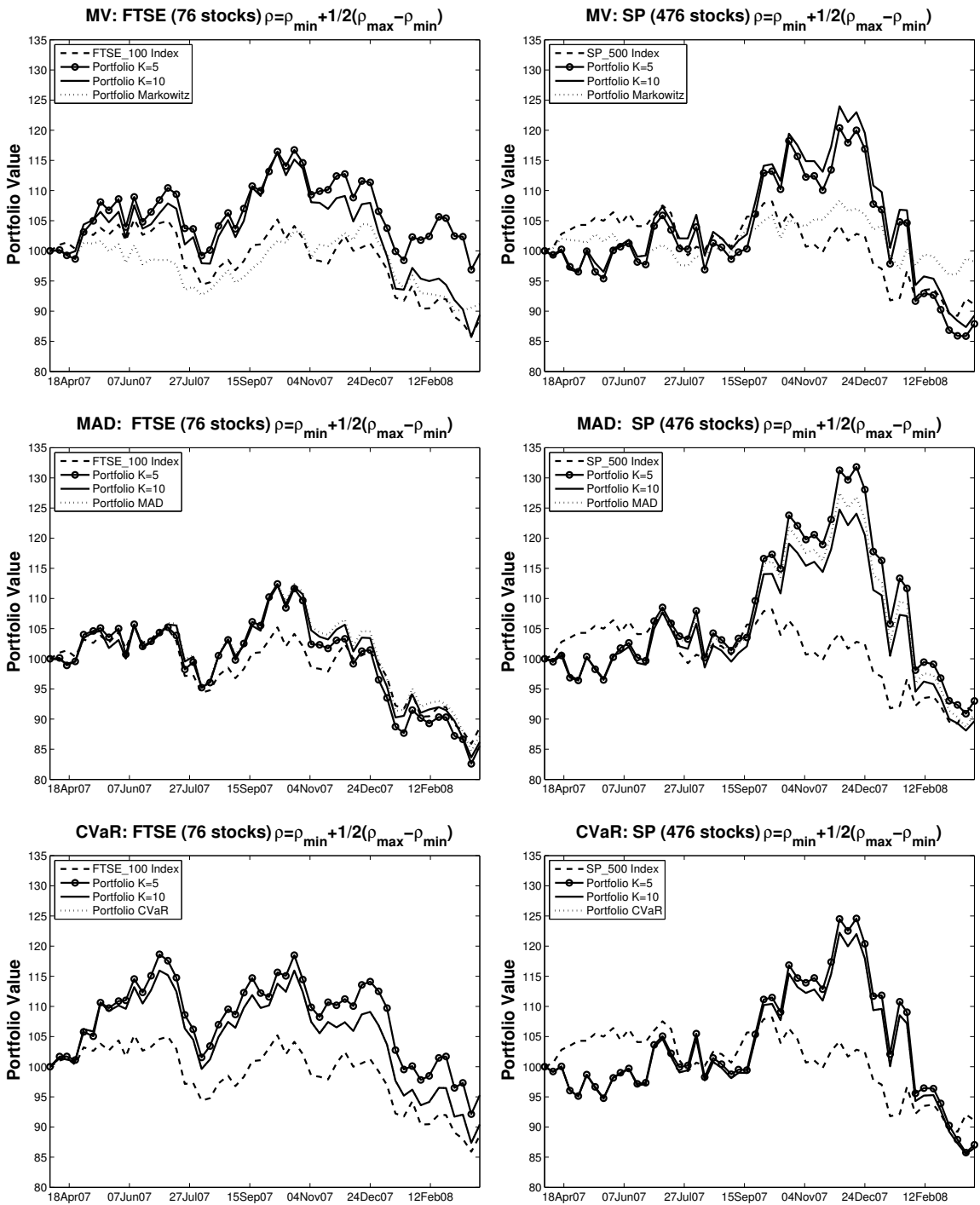

Figure 9: Evaluation of ex-post performances with $\rho = \rho_{min} + 1/2(\rho_{max} - \rho_{min})$ for the Markowitz (top), MAD (middle) and CVaR (bottom) models on the FTSE100 (left) and SP500 (right) data sets

for the more difficult problems.

Finally, a comparison of ex-post performances of the *Limited Asset* models, of the unconstrained models and of the market indices seems to indicate that a strong limitation on the number of assets to hold in the optimal portfolio could generally provide more robust and convenient portfolios.

We plan to improve the computational efficiency of the Increasing Set Algorithm also by exploiting the possibility of parallel computing. Furthermore, we intend to consider quadratic models with additional complex constraints and possibly with different objectives (e.g., index tracking).